\documentclass[10pt,english,journal]{IEEEtran}
\usepackage[T1]{fontenc}
\usepackage[latin9]{inputenc}
\usepackage{geometry}
\usepackage{amsmath}
\usepackage{xpatch}
\usepackage{amssymb}
\usepackage{esint}
\usepackage{mathtools}
\usepackage{amsthm}
\usepackage{nicefrac}
\usepackage{bigints}
\usepackage{mathtools}
\usepackage{blkarray, bigstrut}
\usepackage{physics}
\usepackage{calligra}
\usepackage{graphicx}
\usepackage{epstopdf}
\usepackage{dsfont}
\usepackage{array,ragged2e}
\usepackage{enumitem}
\usepackage{etoolbox}
\usepackage{babel}
\usepackage[nopar]{lipsum}
\usepackage{psfrag}
\usepackage[ruled,linesnumbered]{algorithm2e}
\usepackage{balance}
\usepackage{float}
\usepackage{subfig}
\usepackage{graphicx}
\SetKw{KwBy}{by}

\makeatletter
\newcommand{\removelatexerror}{\let\@latex@error\@gobble}
\makeatother

\xpatchcmd{\proof}{\hskip\labelsep}{\hskip5\labelsep}{}{}  
\makeatletter
\xpatchcmd{\proof}{\@addpunct{.}}{\@addpunct{:}}{}{}
\makeatother

\renewcommand\[{\begin{equation}}
\renewcommand\]{\end{equation}} 
\usepackage{listings}
\usepackage{fancyvrb}
\usepackage{framed}

\usepackage{courier}
\usepackage[usenames,dvipsnames,table]{xcolor}

\definecolor{dkgreen}{rgb}{0,0.3,0}
\definecolor{gray}{rgb}{0.5,0.5,0.5}

\DeclarePairedDelimiter\floor{\lfloor}{\rfloor}

\makeatletter
\newcommand*{\rom}[1]{\expandafter\@slowromancap\romannumeral #1@}
\makeatother

\usepackage{siunitx}
\usepackage{tabu}
\usepackage{booktabs}
\usepackage{multirow}

\usepackage{capt-of}
\usepackage{array}
\usepackage{arydshln}
\begin{document}
\title{A Collaborative Statistical Actor-Critic Learning Approach for 6G Network Slicing Control}
\author{Farhad Rezazadeh$^{1,2}$, Hatim Chergui$^1$, Luis Blanco$^1$, Luis Alonso$^2$, and Christos Verikoukis$^1$\\
{\normalsize{} $^1$ Telecommunications Technological Center of Catalonia (CTTC), Barcelona, Spain\\ $^2$ Technical University of Catalonia (UPC), Barcelona, Spain}\\
{\normalsize{}Contact Emails: \texttt{\{frezazadeh, hchergui, lblanco,  cveri\}@cttc.es, luisg@tsc.upc.edu}}}
\maketitle
\thispagestyle{empty}

\begin{abstract}
Artificial intelligence (AI)-driven zero-touch massive network slicing is envisioned to be a disruptive technology in beyond 5G (B5G)/6G, where tenancy would be extended to the final consumer in the form of advanced digital use-cases. In this paper, we propose a novel model-free deep reinforcement learning (DRL) framework, called \emph{collaborative statistical Actor-Critic} (CS-AC) that enables a scalable and farsighted slice performance management in a 6G-like RAN scenario that is built upon mobile edge computing (MEC) and massive multiple-input multiple-output (mMIMO). In this intent, the proposed CS-AC targets the optimization of the latency cost under a long-term statistical service-level agreement (SLA). In particular, we consider the \emph{Q-th delay percentile} SLA metric and enforce some slice-specific preset constraints on it. Moreover, to implement distributed learners, we propose a developed variant of \emph{soft Actor-Critic} (SAC) with less hyperparameter sensitivity. Finally, we present numerical results to showcase the gain of the adopted approach on our built OpenAI-based network slicing environment and verify the performance in terms of latency, SLA $Q$-th percentile, and time efficiency. To the best of our knowledge, this is the first work that studies the feasibility of an AI-driven approach for massive network slicing under statistical SLA.
\end{abstract}


\section{Introduction}
\IEEEPARstart{Z}{ero-touch} network and service management (ZSM) framework reference architecture \cite{glob1} is an attempt to tackle the new challenges arising from massive and diversified service requirements in the next-generation mobile networks. In this paper, we focus on the algorithmic innovation and solution aspects of the ZSM standard. Network slicing is the embodiment of severing the network into different segments that enables the multiplexing of virtualized and isolated logical networks---or slices---on top of the same physical network infrastructure. This paradigm is a paramount feature in B5G/6G systems that leverages network softwarization and virtualization technologies such as software-defined networking (SDN) and network function virtualization (NFV). Moreover, the SLA guarantees that slice-level quality of service (QoS) is fulfilled by automating the control of underlying performance metrics \cite{globnew02}. To account for the plethora of user patterns over different slices and handle such a heterogeneous and complex network, automated management and orchestration (MANO) operations require a flexible and scalable design that considers also long-term performance. 

This tendency towards fully automated MANO has aroused intensive research interest in the application of AI and DRL to tackle challenging NP-hard tasks. Koo \emph{et al.,} have proposed a DRL-based network slicing method and improved resource utilization and latency performance with time-varying traffic  \cite{glob3}. In \cite{glob4}, the authors have proposed a scheme to effectively allocate network resources. The authors have integrated the alternating direction method of multipliers (ADMM) and DRL where exploit the deep deterministic policy gradient (DDPG) \cite{glob5} as a state-of-the-art (SoA) Actor-Critic technique to learn the optimal policy. Pujol Roig \emph{et al.} have proposed an Actor-Critic, called parameterized action twin (PAT) deterministic policy gradient algorithm where automated MANO allows a central unit (CU) to learn to re-configure resources autonomously \cite{glob6}. Liu \emph{et al.} have studied a new decentralized DRL-based resource orchestration system, to automate dynamic network slicing in wireless edge computing networks \cite{glob7}. In \cite{glob8}, the authors have investigated a demand-aware inter-slice resource management solution based on advantage Actor-Critic (A2C) as a DRL algorithm. In \cite{globenew009}, the authors have proposed two centralized scheduling algorithms that take into account latency and SLA requirements in terms of minimal demand and allocate resources in network slicing systems.

From this SoA overview, it turns out that there is no Actor-Critic-driven  network slicing resource allocation strategy that integrates long-term practical SLA constraints with respect to some key performance indicators (KPIs) while also optimizing service cost. Without loss of generality, this work investigates a multi-tenant network scenario with a developed variant of mMIMO \cite{glob2} and edge computing approach as promising B5G/6G wireless access technologies for optimal resources allocation to minimize the latency and automate the corresponding tasks.  The numerical results show that the proposed distributed resource allocation approach enables SLA enforcement while reducing slice-level average and percentile latency.  Specifically, in this work:
\begin{itemize}
    \item We investigate the feasibility of a multi-objective and multi-action approach where model-free agents learn to jointly allocate optimal power and computing resources to minimize the latency of service provisioning under long-term statistical SLA, namely, \emph{$Q$-th delay percentile}.
    \item We propose a massive DRL-based actor-learner framework dubbed CS-AC. The CS-AC is a software framework for designing and training DRL agents that attempts to address complexity and scalability issues. To cope with control challenges in network slicing such as increased dynamism, heterogeneity, and extended training time of slice instances, we separate the actor from learner where CS-AC can be scaled up to several thousand parallel actors-learners across a large collection of tasks without sacrificing data efficiency. We elaborate the main motivations behind this approach in Sec. \rom{3}.
    \item To implement distributed learners, we combine deep Q-network (DQN) and policy gradients in form of Actor-Critic \cite{glob9} approach and propose a developed variant of SAC \cite{glob10} that we term \emph{stochastic Actor-Critic} to reduce the need for hyperparameter tuning and stabilize the learning procedure.
    \item We develop a 5G RAN network slicing environment called \emph{smartech-v4} where we consider both power and central processor (CPU) resources in a simulator interfaced through OpenAI Gym \cite{glob11}, which is the most famous toolkit in the DRL community.
\end{itemize}

\section{System Model and Problem Formulation}
As shown in Figure 1, we consider a slice-enabling cell-free mMIMO network scheme with edge-cloud computing (central unit/central processing units) capability and distributed access points (APs) connected to a central server via serial fronthaul links. Let us define the cloud radio access network (C-RAN) consisting of $N$ APs that cover $M$ single-antenna users in a downlink setup. The network slicing architecture consist of $L \in \mathbb{N}$ slice instances where each slice accommodates $M_l$ users with $\sum_{l=1}^{L}M_{l} = M$. Each user $M$ sends network slice selection assistance information (NSSAI) to assist the network in selecting a particular network slice where it may be served by a maximum of eight network slices simultaneously \cite{glob12}. We suppose each user just requests one type of service and thereby one slice instance at each decision time step.
\begin{figure}[h!]
\centering
\includegraphics[scale=.45]{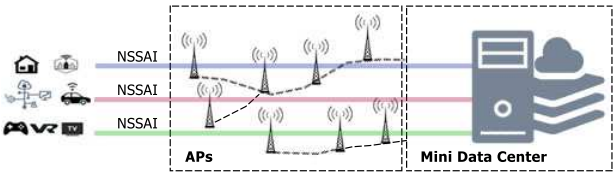}
\caption{The slice-enabling cell-free mMIMO scheme.}
\end{figure}

We consider resource allocation (in NFV) tasks where the mobile network operator (MNO) collects the free and unused resources from the tenants and allocate them to the slices in need. It is done either periodically to avoid over-heading or based on requests of tenants. Moreover, we consider another action concurrently to allocate power to different users. We assume that all the APs are connected to a central server that maintains and deploys a set of virtual network functions (VNFs) to serve the users of distributed APs and also hosts agents for the training process to learn best policies and actions for scaling vertically the computing resources and consequently scale horizontally for VNFs instantiation to minimize the latency according to system states. 

We follow a slotted resource allocation scheme, where the central server allocates resources to new arriving users at the start of the next slot. Indeed, the time horizon is discretized into the decision time step where $t \in \mathbb{N}^+$.
We define \emph{$\jmath_{l}^{(t)}$} as number of new service requests from all APs for $l$-th slice at time step $t$ where it follows an independent and identically distributed Poisson process with parameter $\lambda_{l}^{(t)}$. Therefore, the probability  of new demands to arrive at the central server for time-slot of duration $T$ is given by, $P(\jmath_{l}^{(t)} = \jmath ) = \frac{(\lambda_{l}^{(t)}T)^{\jmath}}{\jmath!}e^{-\lambda_{l}^{(t)}T}$, where $\lambda_{l}^{(t)} = \max\{x \sim \mathcal{N}(\mu_l, \sigma_l, ), 0\}$ is time-varying value to capture slow variations of network traffic over time by sampling  a Gaussian distribution with  parameters $\mu_l$ and $\sigma_l$ \cite{glob6}. Let us define vector of channel gains from the all $N$ APs to the user $m$ as $\mathbf{h}_m = [h_{1,m},h_{2,m}, ..., h_{N,m}]^H \in \mathbb{C}^{N \times 1} $, where $(\cdot)^H$ is the conjugate transpose and $\mathbb{C}$ represents the complex set. Let consider the following channel model \cite{glob13}, $h_{n,m} = 10^{-L^*(d_{n,m})/20}\sqrt{\vartheta_{n,m}\Theta_{n,m}}g_{n,m}$, where $L^*(d_{n,m})$ denotes the path loss with a distance of $d_{n,m}$. Moreover, $\vartheta_{n,m}$ is the antenna gain, $\Theta_{n,m}$ is the shadowing coefficient and $g_{n,m}$ is the small-scale fading coefficient. The beamforming vector $\mathbf{v}_m = [v_{1,m},v_{2,m}, ..., v_{N,m}]^H \in \mathbb{C}^{N \times 1}$ is associated with user $m$ and whose expression is given by \cite{glob14},

\begin{equation}
\mathbf{v}_m = \sqrt{p_m} \frac{\left(\mathbf{I}_N + \sum_{j=1}^{M}\frac{1}{\sigma_{v}^{2}}\mathbf{h}_j\mathbf{h}_j^{H}\right)^{-1}\mathbf{h}_m}{\norm{ \left(\mathbf{I}_N + \sum_{j=1}^{M}\frac{1}{\sigma_{v}^{2}}\mathbf{h}_j\mathbf{h}_j^{H}\right)^{-1}\mathbf{h}_m }},
\end{equation}
where $p_m$ is beamforming power, $\mathbf{I}_N$ denotes the $N\times N$ identity matrix and $\sigma_{v}^{2}$ is the noise variance. Then we define approximate data rate for user $m$ with respect to channel bandwidth $\hat{B}$ and signal-to-interference-plus-noise ratio as follow,
\begin{equation}
R_{m}^{(t)}=\hat{B}\log_2\left(1+\frac{\abs{\mathbf{h}_m^H \mathbf{v}_m}^2}{\sum_{j \neq m}^{M}\abs{\mathbf{h}_m^H \mathbf{v}_j}^2+ \sigma^2 }\right).
\end{equation}

We suppose each user $m$ has a task to be executed. Let define data size of task m as $d_{m}^{(t)} = R_{m}^{(t)} k_{m}^{(t)} $, where $k_{m}^{(t)}$ denotes the transmission time and we consider the coefficient $\zeta_m$ to compute the required computing CPU frequency cycles $\Delta_{m}^{(t)}$ as proportional to data size of corresponding task, $\Delta_{m}^{(t)} = \zeta_m d_{m}^{(t)}$. Then the computing delay is given by,
\begin{equation}
    \mathcal{D}_{m,1}^{(t)} = \frac{\zeta_m d_m}{\mathcal{F}_m},
\end{equation}
where $\mathcal{F}_m$ is computing speed of the edge central server. The $n$-th fronthaul satisfies,
$\chi_{n,m} = \sum_{m \in M}R_m\left[\mathds{1}(\hat{{\chi}}_{n,m} = 1 )\right] \leq   \varphi_{n,th}$ capacity constraint, where $\hat{{\chi}}_{n,m} \in \{0, 1\}$ is a  binary variable to determine the association between the $n$-th AP and the $m$-th user. Let us define $g$-th flow as a competitive flow for $f$-th flow, then the $f$-th flow should wait for transmission of the $g$-th flow. To compute the queuing delay $n$-th link we have,
\begin{equation}
    \mathcal{D}_{n,2}^{(t)} =\frac{\psi\lambda_{\mathcal{D},n}}{ \varphi_{n,th}},
\end{equation}
where $\psi$ denotes the maximum burst size in a fronthaul network \cite{glob16}, and $\lambda_{\mathcal{D},n}$ denotes the number of competitive flows at $n$-th link. Therefore, the total delay for each task $m$ and corresponding fronthaul link $n$ is given by, 
\begin{equation}
    \mathcal{D}_{n,m}^{(t)} = \mathcal{D}_{m,1}^{(t)}+\mathcal{D}_{n,2}^{(t)},
\end{equation}
and thereby the optimization problem defined as,
\begin{subequations}
\label{Prob}
\begin{alignat}{2}
&\!\min        &\qquad& 
\sum_{n=1}^{N}\sum_{m=1}^{M} \mathcal{D}_{n,m}^{(t)}
\\
&\text{subject to} &   
&p_{m}^{(t)} \leq \mathcal{P}_{max},\quad \forall m\in M,\\
&                  &      & \Delta_m^{(t)}\leq \Delta_{th,l},\quad \forall m \in M, \forall l \in L,\\
&                  &      &
\chi_{n,m}^{(t)} \leq  \varphi_{n,th},\quad  \forall n \in N, \forall m \in M\\
&                  &      &f_{Q}^{l}\left(\mathcal{D}_{n,m}^{(1)}, ..., \mathcal{D}_{n,m}^{(t)}\right) \leq \eta_{l}\label{SLA2}, \quad \forall l \in L
\end{alignat}
\end{subequations}
where $\mathcal{P}_{max}$ is the maximum allowable power level and $\{\Delta_{th,l}\}$ is the maximum CPU cycles threshold which can be set based on MNO's preferences and policies.
A typical latency SLA between slice $l$ tenant and the MNO consists on imposing a long-term statistical constraint on the distribution of latency values. In this regard, we invoke the $Q$-th percentile metric $f_{Q}^{(t)}$ that captures, at each step $t$, the actual latency value below which $Q\%$ of latency samples over the measurement interval $i=1,\ldots,t$ are located. We then enforce an slice-specific upper bound $\eta_{l}$ on it. To calculate $f_{Q}^{(t)}(\cdot)$ over set $\mathcal{Y}_{l}=\left\{\mathcal{D}_{n,m}^{(1)}, ..., \mathcal{D}_{n,m}^{(t)}\right\}$, the elements thereof are sorted in the ascending order, i.e., $
\mathcal{Y}_{l}= \{z_1,\ldots,z_{t}\mid z_i < z_{i+1}\},
$
and then the $Q$-th percentile is derived as,
\begin{equation}
f_{Q}^{l}\left(\mathcal{D}_{n,m}^{(1)}, ..., \mathcal{D}_{n,m}^{(t)}\right)= z_j,\,j = \floor*{\frac{Q \times\left(t+1\right)}{100}}.
\end{equation}

In this constrained optimization problem, the system latency closely depends on date rate $R_{m}^{(t)}$, the underlying computing resources delay $\mathcal{D}_{m,1}^{(t)}$, as well as transmission delay $\mathcal{D}_{n,2}^{(t)}$. Specifically, user $m$ allocated power $p_m$  has a direct impact on the achieved data rate and thereby the required computing resources for VNFs as well as the resulting computing delay. This approach presents correlated models where the main aim is to find the best policy for jointly allocating power and computing resources to minimize the service provisioning latency while respecting the long-term statistical SLA. Following model-free DRL-based approaches \cite{glob6}-\cite{glob17}, we formulate the problem from an MDP perspective and develop a new DRL scheme to cope with the underlying high dimensional state and action spaces as detailed in the sequel.

\section{CS-AC}
Figure 2 presents the proposed cross-platform framework (CS-AC) concerning the RAN data center. The architecture consists of six main components: i) The network slicing environment (\emph{smartech-v4}), ii) The slice-level SLA buffer that stores the historical SLA-related metrics, namely, latency states in the current scenario, to instantly calculate their empirical distribution, such as the $Q$-th percentile, and feed it to the DRL block,  iii) The parallel actors typically run on CPUs and interact with network environment to generate new experiences and behaviors ${({s}_t:state,{a}_t:action,{r}_t:reward,{s}_{t+1}:next-{s}_t)}$ asynchronously and enforce the best actions, iv) The experience replay (buffers) to store past experiences while coping with catastrophic interference,  v) The parallel learners to train and optimize the model on GPUs where they sample a random batch $\mathbb{B}_{i}, i \in \mathbb{N}$ for all transitions ${({s}_{t_{B}},{a}_{t_{B}},{r}_{t_{B}},{s}_{t_{B}+1})}$ of $\beta_{i}$, and vi) The memories ($\mathbb{M}_{i}$) for sharing the parameters and models where mitigate the load on learners for model update request and also lessens the average read latency related to the actors. The memories share the policy model of learners ($\phi$) with actors to update their policy ($\mu_{\phi}$). Moreover, the memories can solve the problem of parameter synchronization of actors and learners because they are asynchronous. Indeed, The policy gradient parallelization approach that initially has proposed in A3C \cite{globA3C} can reduce computation time 
and stabilizes the learning.

\begin{figure}[h!]
\centering
\includegraphics[scale=0.41]{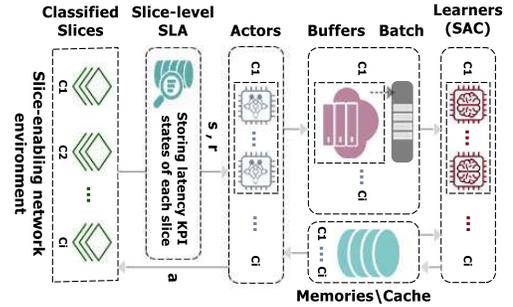}
\caption{The proposed software-based framework for massive network slicing.}
\vspace{-1mm}
\end{figure}

Separating actors from learners in a network slicing is motivated by improving the learning efficiency referring to other high-throughput learning frameworks, such as Gorila \cite{glob19} and Impala \cite{glob20}. Unlike the previous works, the inference model in CS-AC is executed centrally by the learners. This approach can reduce bandwidth requirements for transferring updated model parameters from learners to actors while the multiple buffers and memories mitigate read latency. We can classify slices according to different scenarios and metrics (e.g. QoS, priority, and tenant ID) and assign each class ($c_i$) to a collection of actors and learners. The CS-AC can support unbounded limit actors for a massive network such as network slicing. To reduce waiting time, the CS-AC ignores the slowest actors. Following MDP in RL parlance, the state ($s_t$), action ($a_t$) and proposed reward function ($r_t$) are defined as follows:

\textbf{1) State space:} The state space provides some information about different possible network configurations. Indeed, it helps to learn the best policy (mapping states to actions) through interaction with network slicing parameters. In our scenario, the state transits to the next state at each time step $t$ as input can be characterized by $S^{(t)} = \{S_1^{{(t)}}, S_2^{{(t)}}, S_3^{{(t)}}, S_4^{{(t)}}\}$, where $(S_1^{{(t)}})$ is the number of arrival requests for each slice, $(S_2^{{(t)}})$ is data rate status, $(S_3^{{(t)}})$ refers to computing resources allocated to each slice, and $(S_4^{{(t)}})$ is latency status with respect to latency cost for each slice.

\textbf{2) Action space:} 
We define a continuous multi-action space in telecommunication environment and pursue an experimental approach aiming to allocate power and computing resources to each slice and scrutinize the learning behaviour of the agent in terms of minimizing latency for service provisioning where the allocated power is given by,
\begin{equation}
       \mathcal{A}_{P}^{(t)} \in \{ o  |   o \in \mathbb{R}, 0\leq o \leq \mathcal{P}_{max}^{(t)}\},
\end{equation}
where $\mathcal{P}^{(t)}_{max}$ is an experimental value. Moreover, we consider vertical scaling consists of either scaling up or down, i.e., increasing or decreasing the computing resources, respectively. Therefore, the agents allocate computing resource according to each time step,
\begin{equation}
       \mathcal{A}_{CPU}^{(t)} \in \{ o  |   o \in \mathbb{R},-\sum_{m = 1}^{M}\Delta_{m}^{(t)}\leq o \leq \Delta_{max}   -\sum_{m = 1}^{M}\Delta_{m}^{(t)}\},
\end{equation}
where $\mathcal{A}_{CPU}^{(t)}$ is vertical scaling action for CPU resources. Note that vertical scaling is limited with respect to the amount of free computing resources available $\Delta_{max}$ on the physical server hosting the virtual machine. The complete action space is given by, $\mathcal{A}^{(t)} \triangleq  \mathcal{A}_{CPU}^{(t)} \cup \mathcal{A}_{P}^{(t)}$. Note that we do not consider horizontal scaling and server selection because it requires another algorithm with discrete action space.  

\textbf{3) Reward:} 
The total network cost (Problem 6) is an imprecise and very general metric to guide the agents to learn best policy and thereby select the best actions. To enforce both the statistical and punctual constraints, we introduce the piecewise function ${\Omega}_{l,m}^{(t)}$,
\begin{equation}
    \Omega_{l,m}^{(t)}=-\varrho_{l,m} \mathds{1}\left(\Delta_{m}^{(t)} > \Delta_{th,l} \cup f_{Q}^{l}\left(\mathcal{D}_{n,m}^{(1)}, ..., \mathcal{D}_{n,m}^{(t)}\right) > \eta_{l}\right),
\end{equation}
where $\varrho_{l,m}$ is the penalty coefficient for violating either the CPU constraint or the $Q$-th percentile SLA, which can be fine-tuned. Consequently, the total return is given by,
  \begin{equation}
    r^{(t)} = \frac{1}{\frac{1}{ M^{(t)} }\sum_{n=1}^{N}\sum_{m=1}^{M} \mathcal{D}_{n,m}^{(t)}}+\sum_{l=1}^{L}\sum_{m=1}^{M}{\Omega}_{l,m}^{(t)}.
  \end{equation}
  
We consider the number of users at each decision time step ($M^{(t)}$) to make a balance and normalize the network cost between heavy and low traffic. This return function is used in deep neural network (DNN) training. We propose this reward-penalty technique to increase the expected return (reward, in RL parlance) while minimizing the latency cost.

The learner part of CS-AC uses an Actor-Critic setup based on a developed variant of SAC \cite{glob10}. The DQN and policy gradient are fundamentals of Actor-Critic methods where the actor is a  DNN to parameterize the policy and critic is another DNN to parameterize the value function. Note that the actor task in actor-learner is different from actor task in Actor-Critic method. Unlike the DDPG \cite{glob5} method, the SAC benefits from stochastic policy gradient based on policy gradient theorem \cite{glob23}. The main goal in standard  RL  is to learn a policy $\pi(a_t,s_t)$ to maximize the expected sum of rewards. The SAC method \cite{glob10} benefits from a policy entropy term $\mathcal{H}$. Maximum entropy RL improves the exploration efficiency of the policy. The objective for finite-horizon MDPs is given by, $
    J_{\pi} = \mathbb{E} \left[\sum_{i=t}^{T}\gamma^{i-t}[r_i+\alpha\mathcal{H}(\pi(\cdot|s_i))]  \right]$,
where $\gamma$ is the discount factor and $\alpha$ denotes a temperature parameter to determine the relative importance of the $\mathcal{H}$ against the reward and handle the stochasticity of the optimal policy.


To learn the optimal maximum entropy policies, we use the soft policy iteration method. The convergence and optimality of this approach have been verified in \cite{globe24}. 
Our proposed SAC method incorporates a set of techniques such as double (clipped) Q-learning technique \cite{glob22}, target DNN to compute true target in DQN, the experience replay to memorize past experiences and solve the catastrophic interference issues, and the delayed strategy \cite{glob6} to update the policy, target networks and temperature less frequently than the value network. The goal is to mitigate very high sample complexity and meticulous hyperparameter tuning and also stabilize the learning.

We parameterize functions $Q_{\theta}(s,a)$ and $\pi_{\phi}(a|s)$ to approximate the soft Q-value and policy where $(Q_{\theta_1}, Q_{\theta_2})$ are soft Q-value functions and $(Q_{\theta^{\prime}_1}, Q_{\theta^{\prime}_2})$ are target soft Q-value functions. The updates of $Q_{\theta_1}$,  $Q_{\theta_2}$ based on targets is given by,
\begin{equation}
    y= r+\gamma(\underset{i = 1,2}{\min}Q_{\theta^{\prime}_i}(s^{\prime}, a^{\prime})) - \alpha\log\pi_{\phi}(a^{\prime}|s^{\prime}), \quad a^{\prime}\sim\pi_{\phi}.
\end{equation}
To train the soft Q-value, we can directly minimize,
\begin{equation}
    J_Q(\theta_i) =\mathbb{E} [(y-Q_{\theta_i}(s,a))^2], \quad i=1,2.
\end{equation}

The SAC method leverages a reparameterization trick \cite{globe24} to reduce variance estimates where reparameterize the policy using a neural network transformation $a = f_{\phi}(\xi;s)$. The policy update gradients based on experience replay ($\beta$) is given by, 
\begin{equation}
\begin{aligned}
\nabla_{\phi}J_{\pi}(\phi) =
\mathbb{E} [-\nabla_{\phi}\alpha\log(\pi_{\phi}(a|s))+(\nabla_aQ_{\theta}(s,a)\\-\alpha\nabla_a\log(\pi_{\phi}(a|s))\nabla_{\phi}f_{\phi}(\xi;s))]
\end{aligned}
\end{equation}

The temperature $\alpha$ can be updated through following objective 
$J(\alpha)=\mathbb{E}[-\alpha\log\pi_{\phi}(a|s)-\alpha\mathcal{H}]$. The proposed approach for a single agent  of network slicing is summarized in Algorithm 1 (actor) and Algorithm 2 (Learner).

\begin{algorithm}[h!]
\scriptsize
\SetAlgoLined
Initialize replay buffer $\beta_c$\\
Import network slicing environment (`smartech--v4')\\
Initialize action space $\mathcal{A}$ and state space S\\
t=0\\
\While {t < max\_timesteps}{
  \eIf{t < start\_timesteps}{
   Initial action $a$ = env.action\_space.sample() to fill buffer
   }{
    {Select action using the updated network parameters} $a\sim\mu_{\phi}(a|s)$ \emph{w.r.t.} {Algorithm 2}
  }
  Apply the action in the network slicing\\ Observe next\_state, reward, done, \_ = env.step($a$)\\
  store the new transition ${({s}_t,{a}_t,{r}_t,{s}_{t+1})}$ into $\beta_c$\\
  \If{done}{
   obs, done = env.reset(), False
   }Obtain latest network parameters from $\mathbb{M}_c$ periodically\\t=t+1
 }
\caption{Actor}
\end{algorithm}

\begin{algorithm}[h!]
\scriptsize
\SetAlgoLined
Initialize actor network $\phi$, critic network $\theta$, and temperature $\alpha$\\
Initialize (copy parameters) target networks ${\theta}_1^{\prime}$, ${\theta}_2^{\prime}$\\
Initialize learning rate $\ell_{\mathcal{Z}}, \ell_{\pi},\ell_{\alpha}$\\
Initialize memory $\mathbb{M}_c$
 
 \While {t < max\_timesteps}{
  
   \If{t $\geq$ start\_timesteps}{
   sample batch of transitions ${({s}_{t_{B}},{a}_{t_{B}},{r}_{t_{B}},{s}_{t_{B}+1})}$
   
    $\theta_i\longleftarrow\theta_i-\ell_{Q}\nabla_{\theta_i}J_Q(\theta_i)$, \quad i=1,2 \quad\#Update soft Q-function

   \If{$t \mod freq$}{ 
   $\phi\longleftarrow\phi+\ell_{\pi}\nabla_{\phi}J_{\pi}(\phi)$ \quad \#Update policy weights
   
   $\alpha\longleftarrow\alpha-\ell_{\alpha}\nabla_{\alpha}J(\alpha)$\quad \#Adjust temperature
   
${\theta}_{i}^{\prime}\longleftarrow\tau{\theta}_{i}+(1-\tau){\theta}_{i}^{\prime}$\quad i=1,2\quad \#Update target network
                    }
   Update memory $\mathbb{M}_c$ periodically\\
   Obtain updated parameters from\ $\mathbb{M}_c$ periodically
   
   }
 
 }

\caption{Learner}
\end{algorithm}

\section{Numerical Results}
To evaluate our method (CS-AC) described in section \rom{3}, we use our PyTorch-based custom environment (\emph{smartech-v4}) with a multi-processing approach interfaced through OpenAI Gym \cite{glob11} and compare this method against other SoA DRL approaches, namely, SAC \cite{glob10} and DDPG \cite{glob5}. Note that the benchmarks have a minor change to make algorithms consistent and all of them support continuous action space and state space. We consider a class of slices with three different slices (A, B, and C).  The corresponding class consists of 3 actors, 3 learners, and 2 buffers. We discussed the procedure of generating new service requests in section \rom{2}. Moreover, the size of each task is in the range $[2, 20]$ Mb that is generated uniformly. There exist 10 APs and a maximum of 17 registered subscribers that are assigned randomly to different slices in each decision time step, where the number of subscribers of slice-A is less than slice-B and slice-C. The adopted percentile value during the training is $Q=95\%$.

\begin{table}[h!]
\caption{Network parameters in simulation.}
\scriptsize
\centering
\begin{tabular}{@{}lccccccccc@{}}\toprule
\textbf{Network Parameter} & \textbf{Value}\\ \midrule
\textbf{Channel bandwidth} & 10 MHz\\ \hdashline
\textbf{Background noise ($\sigma^2$)} & -102 dBm\\ \hdashline
\textbf{Antenna gain ($\vartheta_{n,m}$)} & 9 dBi\\ \hdashline
\textbf{Log-normal shadowing ($\Theta_{n,m}$)
} & 8 dB\\ \hdashline
\textbf{Small-scale fading distribution ($g_{n,m}$)} & $\mathcal{C}\mathcal{N}$(0, $I$)\\ \hdashline
\textbf{Path-loss at distance $d_{n,m}$ (km)} & 148.1+37.6 $\log_2$($d_{m,n}$) dB\\ \hdashline
\textbf{Distance $d_{m,n}$} & distributed uniformly [0, 600]\\

\bottomrule
\end{tabular}

\end{table}

Table \rom{1} presents the network parameters. We set hyperparameters of DNNs through extensive experiments \cite{glob25}-\cite{glob26-icc} and adopt a similar architecture for both Actor-Critic and target DNN models. We use $5$ hidden layers and $128$ units per layer with batch size $128$. Unlike SAC and DDPG methods that use ReLU, CS-AC leverages GELU \cite{glob26} for non-linearity. We compute the performance of the algorithms based on the total of $25\times 10^4$ decision time steps and evaluate the average over each $10^4$ iterations with regard to the best $3$ of the $5$ average return episodes. In algorithm 1, $freq = 2$ refers to update interval for updating policy and target soft Q-value networks. Moreover, $\tau = 0.001$ denotes the target smoothing coefficient \cite{glob22}. Note that the curves are smoothed for visual clarity based on confidence interval over $3$ trials. 

In Figure 3, the learning curve of CS-AC outperforms other approaches in the final performance. Indeed, the CS-AC leverages a parallel rollouts technique via following parallel multiple samples or batch gradient descent simultaneously. As we mentioned in section \rom{3}, the reward function benefits from a reward-penalty technique, and agents learn to maximize the average reward while minimizing the constrained latency optimization task (Problem 6).  
\begin{figure}[h!]
\centering
\includegraphics[scale=.57]{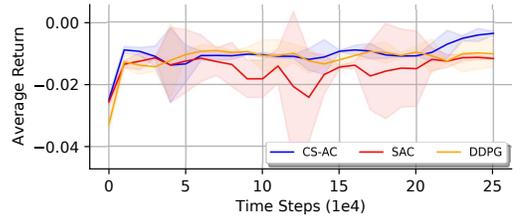}
\caption{Learning curves of smartech-v4 network slicing based on continuous control benchmarks}
\vspace{-1mm}
\end{figure}

As shown in Figure 4, the parallelization approach in CS-AC yields performance improvement significantly compared to SAC and DDPG methods in terms of wall-clock time consumption on the network slicing environment. Note that the Actor-Critic
\begin{figure}[h!]
\centering
\includegraphics[scale=.57]{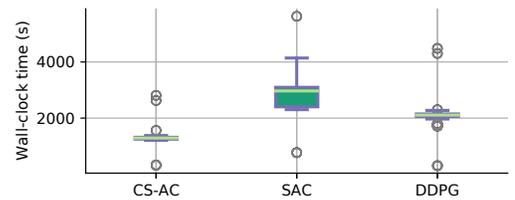}
\caption{Time efficiency comparison.}
\vspace{-1mm}
\end{figure}
\begin{figure*}[t]
\centering
\subfloat[Latency (Slice-A)]{%
      \includegraphics[width=0.21\textwidth]{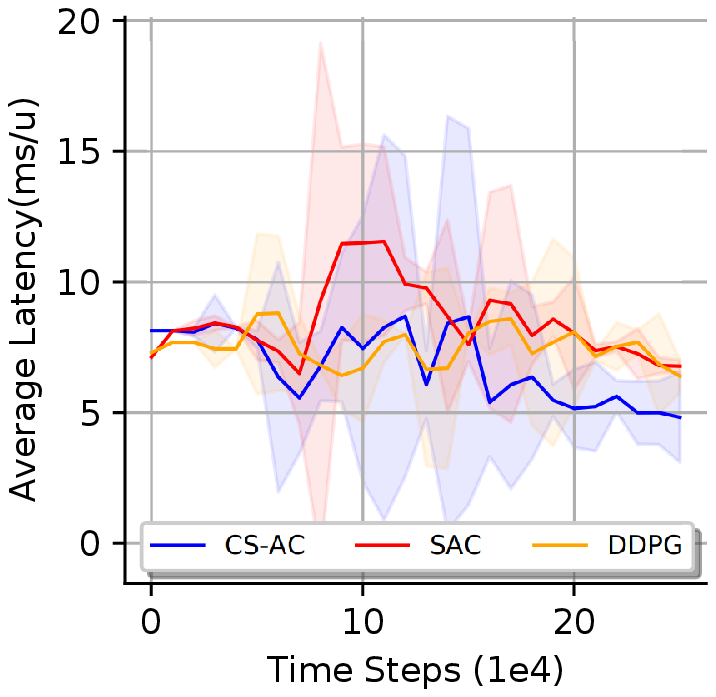}}
\subfloat[Latency (Slice-B)]{%
      \includegraphics[width=0.21\textwidth]{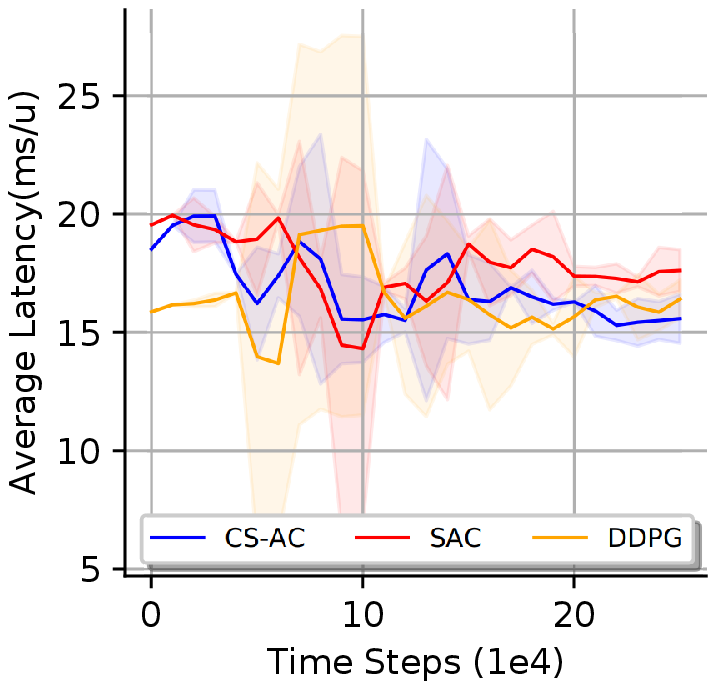}}
\subfloat[Latency (Slice-C)]{%
      \includegraphics[width=0.21\textwidth]{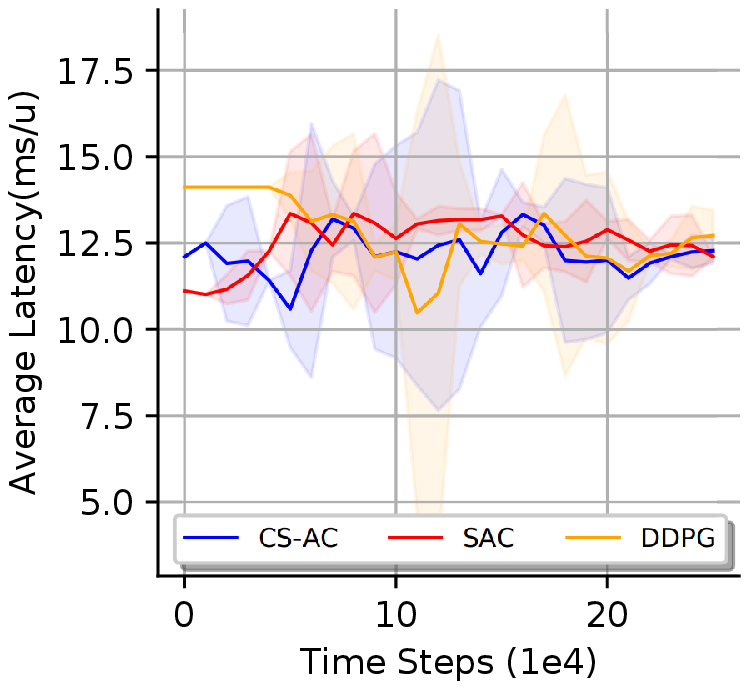}}
\subfloat[Latency percentile in evaluation mode]{%
      \includegraphics[width=0.35\textwidth]{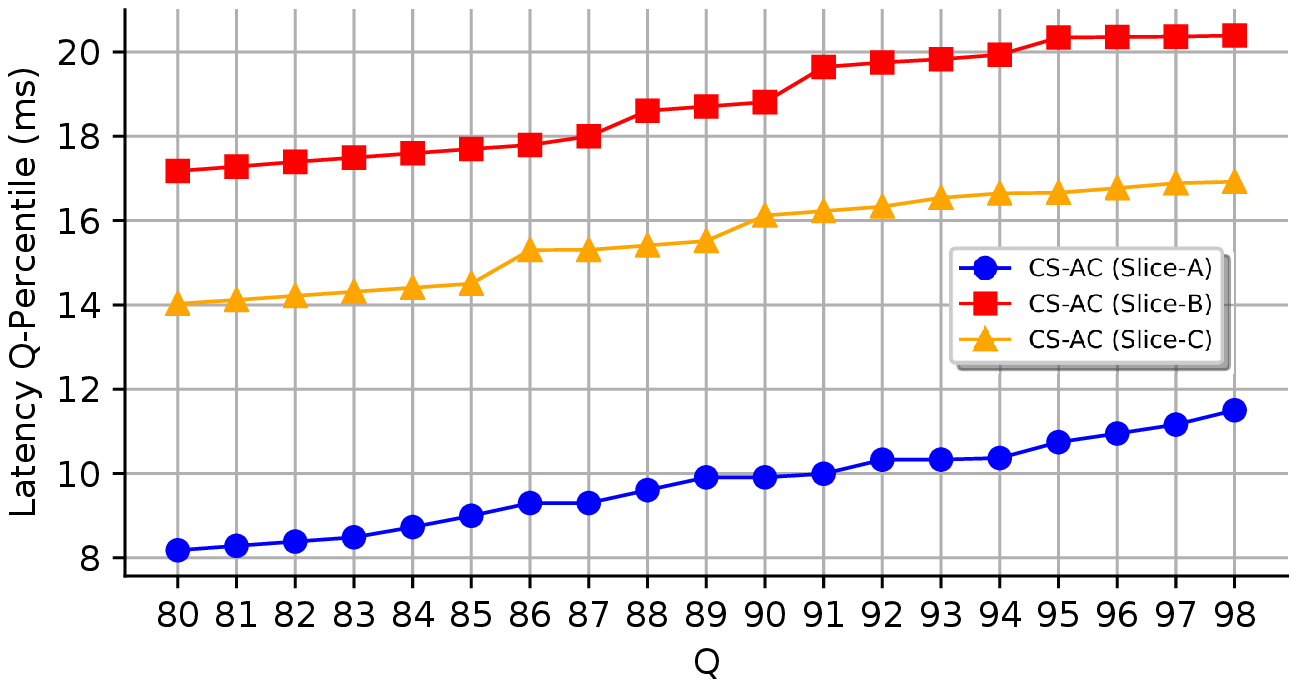}}
      
\caption{Network latency performance for CS-AC,  $\eta = [10, 20, 15]$ ms.}
\vspace{-0.3cm}
\end{figure*}
architecture in CS-AC consists of 5 DNNs while SAC and DDPG have 6 and 4 DNNs respectively and this is the reason for the lower wall-clock time of DDPG compared to SAC. This evaluation was carried out 100 times over averaging of 50 time steps. To analyze the final performance of algorithms, we should consider both average return and wall-clock time.

Figure 5 demonstrates the performance of CS-AC in terms of latency. The agents learn to tune optimal power and computing resources to minimize the latency concerning different traffic demands and network configurations (states). Figures 5-(a), 5-(b), and 5-(c) show that the performance of CS-AS is better than other approaches for slice-A, slice-B, and Slice-C. Indeed, the CS-AS can surmount the curse of dimensionality while coping with the overestimation problem \cite{glob22} in Actor-Critic methods and stabilize the learning procedure. 
Figure 5-(d) presents the latency percentile vs. $Q$ after the training, i.e., in evaluation mode with $\eta = [10, 20, 15]$ ms latency upper thresholds for slice-A, slice-B, and slice-C, respectively. Since the CS-AC has been trained with $95\%$-perectile statistical constraints, we remark that the three slices are approximately respecting the enforced latency upper bound at $Q=95$, i.e. the long-term percentile-based SLA is respected.

\section{Conclusion}
In this paper, we have presented a novel AI-driven software-based framework for control massive network slicing in B5G/6G, dubbed CS-AC. Specifically, we have considered a slice-level statistical DRL method based on the SAC algorithm for allocating power and computing resources dynamically to minimize a latency-aware cost optimization under $Q$-th delay percentile SLA metric. The numerical results of the proposed actor-learner approach have shown that the target $Q$-th percentile is respected while also guaranteeing better performance in terms of latency and time efficiency compared to other SoA DRL benchmarks. As future research directions, we consider addressing distributed resource allocation of end-to-end (E2E) massive slicing in future 6G networks.

\section*{Acknowledgement}This work has been supported in part by the research projects MonB5G (871780), ZEROTO6G, AGAUR (2017-SGR-891), and PROGRESSUS (876868).

\end{document}